\documentclass{article}

\usepackage[preprint,nonatbib]{neurips_2022}

\usepackage[utf8]{inputenc} % allow utf-8 input
\usepackage[T1]{fontenc}    % use 8-bit T1 fonts
\usepackage{hyperref}       % hyperlinks
\usepackage{url}            % simple URL typesetting
\usepackage{booktabs}       % professional-quality tables
\usepackage{amsfonts}       % blackboard math symbols
\usepackage{nicefrac}       % compact symbols for 1/2, etc.
\usepackage{microtype}      % microtypography
\usepackage{xcolor}         % colors
\usepackage{graphicx}
\usepackage{textcomp}
\usepackage{subfigure}

\newcommand{\rsquared}{R$^2$ }

\title{Quantitative Imaging Principles Improves Medical Image Learning}

\author{%
  Lambert T. Leong\thanks{https://github.com/LambertLeong/DXA-VAE} \\
  Department of Molecular Bioscience and Bioengineering\\
  University of Hawaii\\
  \texttt{lambert3@hawaii.edu} \\
\And
  Michael C. Wong \\
  Department of Epidemiology\\
  University of Hawaii Cancer Center\\
  \texttt{mcwong@hawaii.edu} \\
\And
  Yannik Glaser \\
  Department of Computer Science\\
  University of Hawaii\\
  \texttt{yglaser@hawaii.edu} \\
\And
  Thomas Wolfgruber \\
  Department of Epidemiology\\
  University of Hawaii\\
  \texttt{tomwolf@hawaii.edu} \\
\And
  Steven B. Heymsfield \\
  Pennington Biomedical Research Center\\
  Louisiana State University\\
  \texttt{Steven.Heymsfield@pbrc.edu} \\
\And
  Peter Sadowski \\
  Department of Computer Science\\
  University of Hawaii\\
  \texttt{peter.sadowski@hawaii.edu} \\
\And
  John A. Shepherd \\
  Department of Epidemiology\\
  University of Hawaii Cancer Center\\
  \texttt{johnshep@hawaii.edu} \\
} 
\begin{document}

\maketitle

\begin{abstract}
Fundamental differences between natural and medical images have recently favored the use of self-supervised learning (SSL) over ImageNet transfer learning for medical image applications. Differences between image types are primarily due to the imaging modality and medical images utilize a wide range of physics based techniques while natural images are captured using only visible light. While many have demonstrated that SSL on medical images has resulted in better downstream task performance, our work suggests that more performance can be gained. The scientific principles which are used to acquire medical images are not often considered when constructing learning problems. For this reason, we propose incorporating quantitative imaging principles during generative SSL to improve image quality and quantitative biological accuracy. We show that this training schema results in better starting states for downstream supervised training on limited data. Our model also generates images that validate on clinical quantitative analysis software.
\end{abstract}
 
\section{Introduction}

Deep learning is a promising methodology for medical imaging applications. However, acquiring large labeled medical training datasets needed to train robust models is difficult and major barriers result from necessary laws that protect patient privacy. Transfer learning has been the most common method used to address data scarcity in which models would first leverage ImageNet~\cite{deng_imagenet_2009}, for instance, and subsequently be fine-tuned with the limited medical domain data. While some have reported benefits from transfer learning~\cite{CHEPLYGINA2019} its efficiency and efficacy for medical image learning problems is questionable~\cite{raghu_transfusion_2019}. Medical images are inherently different from natural images contained in transfer learning datasets like ImageNet. For example, subtle textures, pixel intensity, scale, and local structures are significant in medical images~\cite{holmberg_self-supervised_2020, wang_chestx-ray_2019} while large-scale shapes are of relevance in natural images where a global subject is often present. Stark differences between natural and medical images highlight the need for improved domain specific methods.

Curation efforts of larger public medical imaging datasets~\cite{rajpurkar_chexnet_2017, sudlow_uk_2015, gulshan_2016,mura,mrnet} have made unsupervised and self-supervised learning (SSL) a more promising alternative method than natural image transfer learning alone. SSL enables the leveraging of unlabeled image data from the medical domain and often results in better starting states for downstream task-specific tuning with limited data~\cite{azizi_big_2021}. Despite performance improvement as a result of SSL, most medical imaging SSL implementations mirror implementations used for natural images thus not considering the differences in medical images. In other words, medical image learning problems are often framed similar to learning problems involving natural images and use similar tools, network architectures, or loss functions. Consequently, medical image model learning may be hindered by the way in which some medical images, as a data type, are misperceived and the method in which learning tasks are constructed. 

Images from modalities which capture internal body components such as X-ray, computed tomography (CT), or magnetic resonance imaging (MRI) are often viewed and used by model developers in a qualitative way similar to the way natural images are used. However, many medical  image types should be used as quantitative maps instead of just qualitative anatomical images~\cite{gulani_quantitative_2020}. Internal imaging modalities rely on specific biophysical interactions to generate images in which each pixel corresponds to a specific biological, physical, and physiological property~\cite{bushberg_essential_2002}. This is especially relevant to the case of image reconstruction learning task which includes generative self-supervised learning and cycle generative adversarial networks (GANs)~\cite{CycleGAN2017}. Often, image quality metrics such as peak signal to noise ratios (PSNR) or structural similarity indexes (SSIM)~\cite{hore_2010} are reported for reconstructions of medical images yet the accuracy as it relates quantitative imaging measurements or biology is underexplored. In fact, the lack of quantitative accuracy has limited the adoption of deep learning generated images into clinical practice since it fails to validate quantitatively on commercial clinical systems and software. 

In this work, we explore an SSL training method to incorporate quantitative medical imaging principles with the main objective of improving biological accuracy for image reconstruction and subsequent learning tasks. This work focuses on the dual energy X-ray absorptiometry (DXA) image type to illustrate the significance of our contributions which are as follows:

\begin{enumerate}
  \item Using quantitative imaging knowledge, we constructed a loss function to constrain the SSL of a variational auto-encoder (VAE) which resulted in improved image reconstruction quality as well as improved quantitative biological accuracy.
  \item The SSL of the VAE with the custom loss function results in a trained encoder subnetwork which outperformed an ImageNet transfer learning model on a medical imaging regression task.
  \item The trained VAE generator subnetwork resulting from SSL demonstrated superior performance on a cross modality image translation task compared to a model not utilizing quantitative medical imaging knowledge during pretraining. 
\end{enumerate}

\section{Related Work}
We provide an overview of prior work in medical imaging pretraining methods as well as works on quantitative DXA image deep learning.

\paragraph{Pretraining}strategies are employed in medical image deep learning to overcome a lack of available training images, labels, or both. Reviews have highlighted the major differences between medical and non-medical imaging datasets and conclude that the benefits of natural image (i.e. ImageNet) transfer learning are not clear~\cite{azizi_big_2021,morid_scoping_2021,litjens_survey_2017}.  Recent focus has turned to self-supervised methods which include methods such as contrastive learning~\cite{he_momentum_2020,chen_big_2020}, auto-encoding~\cite{vincent_extracting_2008,bengio_2013}, and adversarial~\cite{ledig_2016,donahue_2019,ARMANIOUS_2020} learning as alternatives strategies. The aforementioned methods allow models to leverage more domain specific data, i.e. medical images, and these SSL methods have led to improved image analysis, segmentation, and classification~\cite{holmberg_self-supervised_2020,chen_self-supervised_2019,ghesu_2022,chaitanya2020contrastive,tajbakhsh2019surrogate}. Generative image tasks such as denoising, image reconstruction, and artifact removal~\cite{he_autoencoder_2021,matzkin_self-supervised_2020} have also benefited from SSL methods such as multimodal imaging models and unpaired data learning by means of cycleGANs~\cite{wang_unsupervised_2018,shan2019competitive,kong_breaking_2021,ma_cycle_2020,cao_auto-gan_2020}. This work builds upon the notion that medical images are unique to images used in common transfer learning datasets and that SSL provides better results in performance for downstream medical imaging tasks~\cite{chen_big_2020,chowdhury2021applying}. This work differs from previous works because our generative SSL method utilizes the knowledge of medical imaging physics~\cite{bushberg_essential_2002} and model generated images are evaluated on a quantitative biological imaging basis and not just for image quality~\cite{hore_2010,shin2018medical,ARMANIOUS_2020}.

\paragraph{DXA}is a quantitative imaging modality and has long been considered the criterion method for measuring body composition~\cite{shepherd2017body} in addition to its primary and popular use of measuring bone density and quality for osteoporosis~\cite{blake_role_2007}. Deep learning DXA modeling primarily focuses on predicting bone density metrics from images~\cite{xiao_prediction_2020,zhang_deep_2020,krishnaraj_simulating_2019} and work on body composition evaluation or DXA image reconstruction or generation is limited~\cite{yoo_deep_2022}. Clinical software is used to derive body composition from DXA images and while work has been done to reconstruct high quality DXA images with deep learning~\cite{wang_pixel-wise_2021}, none have performed quantitative body composition analysis on such images. Generating quantitatively accurate images is a more difficult problem and we turn to informed deep learning methods to address this problem. Domain informed learning such as physics, biology, or etc.~\cite{raissi_physics-informed_2019,yazdani_systems_2020}, has shown to significantly improve model learning and performance. These works demonstrate the benefit of further constraining the learning of models to the concepts, principles, and law of the domain for which the task it is being trained to accomplish~\cite{takeishi2021physics,oh2020cycleqsm}. 
\section{Approach}
\label{sec:approach}

We first detail the DXA-only dataset used for SSL and the paired dataset used to compare subnetworks trained with and without a custom DXA loss function to networks utilizing ImageNet transfer learning. We then describe the SSL model architecture, followed by our custom DXA loss function. Last, we describe how quantitative biological body composition is obtained from our actual and predicted DXA scans. 

\subsection{Datasets}
\label{sec:dataset}
Data used for modeling and experiments were curated from multiple National Institutes of Health (NIH) studies. All participants gave informed consent, and the study protocols were approved by the Institutional Review Board (IRB) for each respective study. All DXA images used for modeling, testing, and subsequent experiments were acquired on Hologic DXA systems (Hologic Inc., MA, USA) and used in the raw image format. The Fit3D Proscanner (Fit3D Inc., CA, USA) was used to acquire 3D body scans which were then standardized to the same pose and number of vertices with the Meshcapade API (Meshcapade GmbH, Tübingen, Germany).

The SSL VAE was trained and evaluated using 16,002 whole-body DXA scans acquired as a part of the Health ABC Study~\cite{habc_2,newman2003strength} and Bone Mineral Density in Childhood Study~\cite{bmdcs,kalkwarf2007bone}. Data was split into a train, validation, and holdout test set using an 80, 10, 10\% split which was stratified by patient age. Splits were also performed by patient key to avoid data leakage. Therefore duplicate scans of the same patient, if any, remained together in the same split.

Experiments conducted to validate our SSL VAE network utilized two additional datasets that had no overlap with those used to train the VAE. Paired images of DXA and 3D body surface scans were acquired as a part of the Shape Up! Adults (R01DK109008) and Shape Up! Kids (R01DK111698) studies. These scans were used to evaluate the trained VAE encoder and generator in two separate experiments. The 3D body surface scan yielded many 3D anthropometric measurements such as waist circumference, body volume, and torso length, to name a few. The combined dataset consisted of 1107 unique DXA/3D scan pairs which were split by patient key into a train, validation, and holdout test set using an 80, 10, 10\% split.

\subsection{Self-supervised Variational Auto-Encoder (VAE)}

The decision to use a VAE was due to the modularity of the architecture and VAEs are commonly used for SSL generative reconstruction~\cite{vincent_extracting_2008}. For instance, the VAE consists of two main subnetwork components which include the encoder and the decoder or generator. We used the trained subcomponents to investigate possible improved starting points for training with our small paired dataset. The Densenet121~\cite{densenet} architecture was used as the encoder with a single convolutional layer prepended to handle the six channel DXA input. The generator consisted of consecutive units of bilinear upsampling and 2D convolutional layers. 

\subsection{DXA Loss Function}
\label{sec:dxa_loss}
Raw DXA images from this particular manufacturer are acquired by cycling between two X-rays energies, low and high, and three filters, air, tissue, and bone, which results in a six channel image with a pixel depth of 14-bits. The search space of possible pixel values is large and therefore, we exploit known relationships between the raw DXA channels to constrain the pixel level predictions of the VAE for improved learned reconstructions. 

Different attenuation values are obtained when X-ray photons at both low and high energies pass through the same material; a body in this case~\cite{mazess1970determining,mazess1990dual,gotfredsen1986measurement}. The observed attenuation at low energy can be expressed as a ratio (R) to the observed attenuation at high energy~\cite{pietrobelli1996dual}. The voltages used to generate the X-rays on a DXA system are predetermined and fixed and thus, extensive studies have been performed to map chemical elements to corresponding R values~\cite{white1980measured,hubbell1982photon,hubbell1969photon}. R values for molecular components such as fatty acids, protein, and triglycerides found in soft tissue as well as calcium hydroxyapatite composing bone are also well characterized~\cite{richmond_task_1985}. As a result, there are theoretically-derived R values that are specific to biological composition and this knowledge can be used to constrain possible pixel values.

Every pixel on all DXA image channels represent X-ray attenuations at that given point in 2D space~\cite{shepherd2001determining}. Therefore, the R value can be calculated at each pixel location by dividing the low energy image channels by the high energy image channels to obtain R images or quantitative maps of varying fractions of biological molecules. Our DXA loss function evaluates both the reconstruction of the raw six channel DXA images (low and high energy images plus 4 additional calibration images) and the R images generated from each reconstruction. Using a perceptual loss [61], feature maps are obtained for actual and predicted raw DXA and R images using a pretrained VGG-16 with frozen weights. The mean absolute error (MAE) is computed for both the raw DXA image and the R image feature maps and the DXA loss is the sum of that MAE for both image types.

\subsection{Quantitative DXA Analysis for Body Composition Evaluation}
\label{sec:apex}
In clinical practice, commercial software is used to analyze DXA images to obtain measures of body composition for the entire body as well as for standardized and adhoc-defined subregions. Compositional analysis includes the quantification of fat, lean, and bone mass as well as bone mineral density (BMD) and bone mineral content (BMC). In this work, Hologic’s Apex 5.4 software was used to assess biological accuracy, with respect to body composition, of deep learning generated DXA images.  This software is meant to be used in a clinical setting and not designed for high throughput analysis. Additionally, the automated analysis features still requires trained personnel to assess that regions of interest segmented the anatomical landmarks properly. Failure to properly identify regions of interest will result in inaccurate body composition.  This manual component limited the number of images that could be analyzed quantitatively. Qualitative metrics were reported on the entire testset for all experiments however, results tables for quantitative body composition comparisons only report on 117 images.

\section{Experiments}
\label{sec:expirements}
\begin{figure}[h]
  \centering
  \includegraphics[width=.58\textwidth]{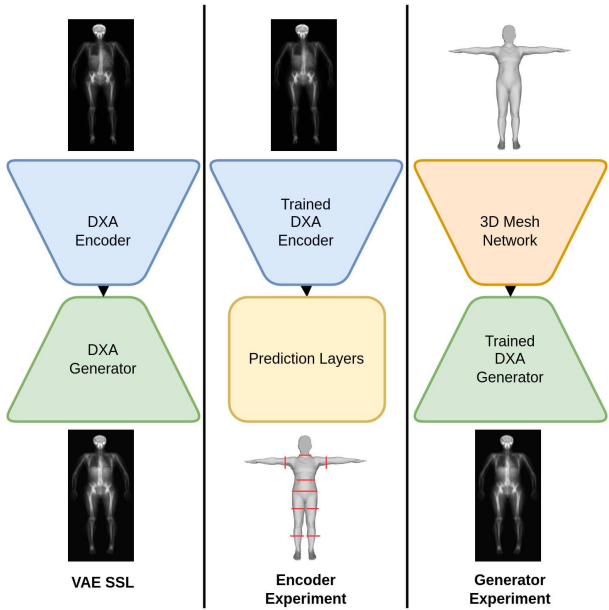}
  \caption{Training setup for SSL of the VAE model on unpaired DXA scans (left), pretraining encoder experimental setup to predict 3D anthropometry from  DXA scans (middle), and generator pretraining experimental setup to predict DXA scans from 3D scans (right).}
  \label{fig:networks}
\end{figure}

In this section we first compare image reconstructions from VAE models trained with and without the DXA loss function. We then conduct two experiments to show that utilizing trained VAE subnetworks results in better model performance when fine-tuned for specific tasks on smaller datasets, See Figure~\ref{fig:networks}. Hyperparameters for all following models were tuned using the SHERPA~\cite{sherpa} python module and all modeling was performed on NVIDIA DGX-1 tesla v100 GPUs.

\subsection{SSL of VAE with DXA Loss}
\label{sec:exp_vae}

The primary motivation for incorporating the DXA loss function was to improve biological accuracy of pixel values by enforcing specific relationships between image channels. Using the same dataset and data split, two identical VAEs were trained starting from randomly initialized weights. The noDXA-VAE was trained using only a perceptual loss while the DXA-VAE utilized the DXA loss detailed in Section\ref{sec:dxa_loss}. We compared the quality of the image reconstructions by computing the PSNR, SSIM, and normalized mean absolute error (NMAE) for both the actual and reconstructed images.  Image quality metrics were computed for images resulting from both the noDXA-VAE and the DXA-VAE and the average values obtained by the test set are shown in Table~\ref{tab:vae-cv}.

\begin{table}[h!]
  \caption{Image quality comparisons of DXA image predicted by self-supervised models when trained without (noDXA-VAE) and with (DXA-VAE) the DXA loss functions.}
  \label{tab:vae-cv}
  \centering
\resizebox{.45\textwidth}{!}{
\begin{tabular}{l|rrr}
\toprule
Model & \multicolumn{1}{l}{PSNR \textuparrow} & \multicolumn{1}{l}{SSIM \textuparrow} & \multicolumn{1}{l}{NMAE \textdownarrow} \\ 
%Model & \multicolumn{1}{l}{PSNR \textuparrow} & \multicolumn{1}{l}{SSIM %\textuparrow} & \multicolumn{1}{l}{NMAE \textdownarrow}\\ 
\hline
%\midrule
noDXA-VAE & 49.52 & 0.997 & 0.038 \\
DXA-VAE & \textbf{51.15} & \textbf{0.998} & \textbf{0.032} \\
\bottomrule
\end{tabular}
}
\end{table}

The DXA-VAE model resulted in better PSNR, SSIM, and NMAE of 51.15, 0.998,	and 0.032, respectively when compared to the noDXA-VAE metrics of 49.52, 0.997, and 0.038, respectively, shown in Table~\ref{tab:vae-cv}. These results suggest that the DXA loss improves the generation of DXA image however, the quality metrics indicate only a slight improvement. 

\begin{figure}[h!]
  \centering
  \includegraphics[width=.64\textwidth]{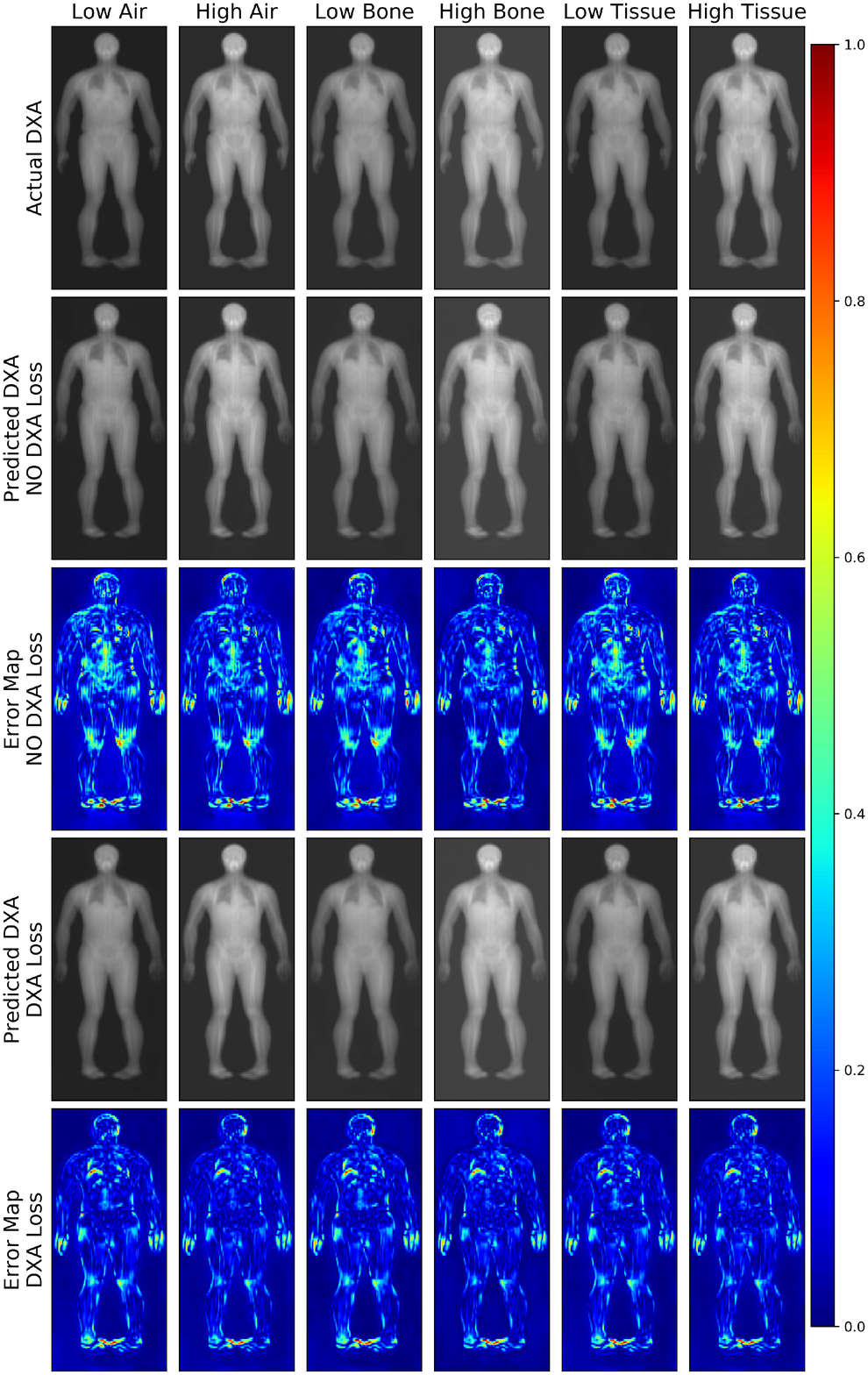}
  %\includegraphics[width=1\textwidth]{vae_dxaloss_compare_sideby.png}
  %\fbox{\rule[-.5cm]{0cm}{4cm} \rule[-.5cm]{4cm}{0cm}}
  \caption{Predicted DXA scan imaging channels and corresponding error maps (represented as percentages) from VAE models strained without and with the DXA loss function}
  \label{fig:vae_errors}
\end{figure}

Figure~\ref{fig:vae_errors} shows representative results test images generated with and without training with the DXA loss. Each column represents one of the six raw DXA channels which. The first row contains the real DXA scan, the second and fourth rows contain a predicted DXA from each SSL VAE training scenario (no DXA loss and DXA loss), and the third and fifth rows contain the error maps computed from the percent difference between the actual and predicted images. Error maps indicate more difference in the prediction from the noDXA-VAE when compared to the network which utilized the DXA loss.

To fully understand the benefit of training with the DXA loss, we evaluate the biological composition of the predicted image using clinical software. We report quantifications of lean and fat tissue as well as bone for the entire body as well as subregions which include just the arms and just the legs alone. Compositional measurements are reported in Table~\ref{tab:vae-apex}.

\begin{table}[h!]
  \caption{Quantitative body composition analysis comparing images predicted from the noDXA-VAE and DXA-VAE.}
  \label{tab:vae-apex}
  \centering
\resizebox{.69\textwidth}{!}{
\begin{tabular}{lrrrr}
\toprule
Measurements & \multicolumn{1}{l}{\begin{tabular}[c]{@{}l@{}}\rsquared\\ noDXA-VAE\end{tabular}} & \multicolumn{1}{l}{\begin{tabular}[c]{@{}l@{}}R$^2$\\ DXA-VAE\end{tabular}} & \multicolumn{1}{l}{\begin{tabular}[c]{@{}l@{}}RMSE\\ noDXA-VAE\end{tabular}} & \multicolumn{1}{l}{\begin{tabular}[c]{@{}l@{}}RMSE\\ DXA-VAE\end{tabular}} \\
\midrule
\midrule
Arm BMC & 0.61 & \textbf{0.88} & 0.04 & 0.02 \\
Arm BMD & 0.49 & \textbf{0.84} & 0.09 & 0.07 \\
Arm Fat & 0.22 & \textbf{0.63} & 0.66 & 0.64 \\
Arm Lean & 0.68 & \textbf{0.89} & 0.73 & 0.57 \\
Leg BMC & 0.67 & \textbf{0.87} & 0.06 & 0.04 \\
Leg BMD & 0.47 & \textbf{0.76} & 0.08 & 0.03 \\
Leg Fat & 0.31 & \textbf{0.70} & 1.43 & 1.51 \\
Leg Lean & 0.68 & \textbf{0.88} & 1.68 & 1.39 \\
Total BMC & 0.63 & \textbf{0.89} & 0.34 & 0.21 \\
Total BMD & 0.34 & \textbf{0.75} & 0.10 & 0.05 \\
Total Fat & 0.42 & \textbf{0.80} & 6.84 & 7.88 \\
Total Lean & 0.73 & \textbf{0.91} & 8.95 & 8.46 \\
Total Mass & 0.79 & \textbf{0.99} & 9.82 & 1.43 \\
\bottomrule
\end{tabular}
}
\end{table}

Despite only a modest increase with respect to image quality, training with the DXA loss results in images with more accurate biological composition. DXA-VAE images resulted in better quantitative accuracy as captured by higher \rsquared values for all composition measurements. Using the DXA loss resulted in an overall better model and reconstructed images were of higher quality and more accurate with respect to body composition. The DXA-VAE which was trained in a SSL fashion was then used to investigate the effectiveness of pretraining using domain specific data, i.e. medical images.

\subsection{SSL VAE Encoder Performance Evaluation}

While DXA scans are 2D image types, the pixel values are related to X-ray attenuation through varying thickness of soft tissue and bone. As such, the objective of this modeling was to predict 3D anthropometry from DXA scans. This dataset consisted of DXA scans and corresponding anthropometrics measurements which include circumferences, volumes, and surface areas of various parts of the body. We trained three models, identical in architecture, that differed only in their pretraining methodology. One model utilized a common transfer learning strategy by loading the Densenet121 portion of the network with ImageNet weights while the other two models were created from the encoder portion of our trained DXA-VAE, one trained without the DXA loss function and the other trained with. Hyperparameters, architectures, and data splits were kept consistent across all models to ensure fair comparisons. Prediction results from the holdout test set are shown in Table~\ref{tab:instant_anthro}.

\begin{table}[h!]
  \caption{Comparisons of ImageNet transfer learning, SSL of VAE without the DXA loss function, and SSL of VAE with the DXA loss function for predicting 3D anthropometry from DXA scans.}
  \label{tab:instant_anthro}
  \centering
\resizebox{.999\textwidth}{!}{
\begin{tabular}{llrrrrrr}
\toprule
Measurment & \begin{tabular}[c]{@{}l@{}}Anatomical \\ Location\end{tabular} & \multicolumn{1}{l}{\begin{tabular}[c]{@{}l@{}}\rsquared\\ ImageNet\end{tabular}} & \multicolumn{1}{l}{\begin{tabular}[c]{@{}l@{}}\rsquared\\ noDXA-VAE\\ Encoder\end{tabular}} & \multicolumn{1}{l}{\begin{tabular}[c]{@{}l@{}}\rsquared\\ DXA-VAE\\ Encoder\end{tabular}} & \multicolumn{1}{l}{\begin{tabular}[c]{@{}l@{}}RMSE\\ ImageNet\end{tabular}} & \multicolumn{1}{l}{\begin{tabular}[c]{@{}l@{}}RMSE\\ noDXA-VAE\\ Encoder\end{tabular}} & \multicolumn{1}{l}{\begin{tabular}[c]{@{}l@{}}RMSE\\ DXA-VAE\\ Encoder\end{tabular}} \\
\midrule
\midrule
Circumference (cm) & Bicep Left & 0.84 & 0.85 & \textbf{0.88} & 2.00 & 1.88 & 1.63 \\
Circumference (cm) & Bicep Right & 0.86 & 0.88 & \textbf{0.94} & 1.80 & 1.63 & 1.22 \\
Circumference (cm) & Calf Left & 0.72 & 0.77 & \textbf{0.85} & 1.75 & 1.51 & 1.26 \\
Circumference (cm) & Calf Right & 0.76 & 0.80 & \textbf{0.88} & 1.65 & 1.44 & 1.17 \\
Circumference (cm) & Hip & 0.88 & 0.89 & \textbf{0.93} & 3.80 & 3.69 & 3.17 \\
Circumference (cm) & Thigh Left & 0.78 & 0.80 & \textbf{0.82} & 2.18 & 2.12 & 2.01 \\
Circumference (cm) & Thigh Right & 0.79 & 0.81 & \textbf{0.85} & 2.26 & 2.12 & 1.99 \\
Circumference (cm) & Waist & 0.92 & 0.92 & \textbf{0.94} & 3.53 & 3.51 & 3.40 \\
Surface Area (cm$^2$) & Whole Body & 0.96 & 0.96 & 0.96 & 0.40 & 0.39 & 0.39 \\
Volume (L) & Arm Left & 0.82 & 0.83 & \textbf{0.86} & 0.53 & 0.50 & 0.39 \\
Volume (L) & Arm Right & 0.86 & 0.86 & \textbf{0.91} & 0.40 & 0.39 & 0.29 \\
Volume (L) & Leg Left & 0.67 & 0.71 & \textbf{0.76} & 0.78 & 0.74 & 0.70 \\
Volume (L) & Leg Right & 0.70 & 0.75 & \textbf{0.78} & 0.78 & 0.71 & 0.67 \\
Volume (L) & Torso Volume & 0.94 & 0.94 & \textbf{0.96} & 0.35 & 0.34 & 0.27 \\
Volume (L) & Total  Volume & 0.97 & 0.97 & 0.97 & 0.31 & 0.31 & 0.30 \\
\bottomrule
\end{tabular}
}
\end{table}

The results in Table~\ref{tab:instant_anthro} indicate that transfer learning with ImageNet may be a reasonable approach for some learning problems involving medical imaging. However, the results also indicate that SSL on domain specific data will yield superior performance when compared to transfer learning from natural images. SSL VAE encoder models trained with the DXA loss resulted in the best \rsquared for all anthropometry measurements with the exception of total volume and total surface area. All models achieved similar performance \rsquared of 0.97 and 0.96, respectively. The best anthropometry predictions (bolded in Table~\ref{tab:instant_anthro}) resulted from the model which included the VAE encoder pretrained with the DXA loss function. 

\subsection{SSL VAE Generator Performance Evaluation}
\label{sec:pseudo-dxa}
The motivation for this experiment was to explore the notion that one's exterior 3D body shape is directly related to the underlying bone and soft tissue distribution or composition. Here we again used a paired dataset consisting of 3D body scans and DXA scans from the same individual taken at the same time point. The objective was to derive a mapping to generate DXA scans for a given 3D scan. Just as in the previous section, we are comparing performance improvements brought about by SSL which incorporated quantitative imaging physics. In this experiment, however, we used the trained VAE decoder or generator instead of the encoder subnetwork. The architecture of the 3D scan or mesh network was modeled after the Pointnet++~\cite{qi_pointnet_2017} model and it mapped to the latent space of the DXA generator. For fair comparison, a second model was trained using the same hyperparameters, data split, and architecture with the only difference being that the generator resulted from VAE SSL without the DXA loss function. We evaluate both 3D to DXA or Pseudo-DXA models on the basis of image quality (Table~\ref{tab:pseudo-cv}) and biological composition (Table~\ref{tab:pseudo-apex}).

\begin{table}[h!]
  \caption{Image quality comparisons of DXA image predicted from 3D body surface by networks generators trained without (noDXA-VAE) and with (DXA-VAE) the DXA loss functions.}
  \label{tab:pseudo-cv}
  \centering
\resizebox{.55\textwidth}{!}{
\begin{tabular}{l|rrr}
\toprule
Model & \multicolumn{1}{l}{PSNR \textuparrow} & \multicolumn{1}{l}{SSIM \textuparrow} & \multicolumn{1}{l}{NMAE \textdownarrow} \\ \hline

noDXA-VAE Generator & 35.14 & 0.897 & 0.201 \\
DXA-VAE Generator & \textbf{38.49} & \textbf{0.938} & \textbf{0.130} \\ 
\bottomrule
\end{tabular}
}
\end{table}

\begin{figure}[h]
    \centering
    \subfigure[Example of a participant 3D body surface scan.]{\label{fig:3dscan}
    \includegraphics[width=.33\textwidth]{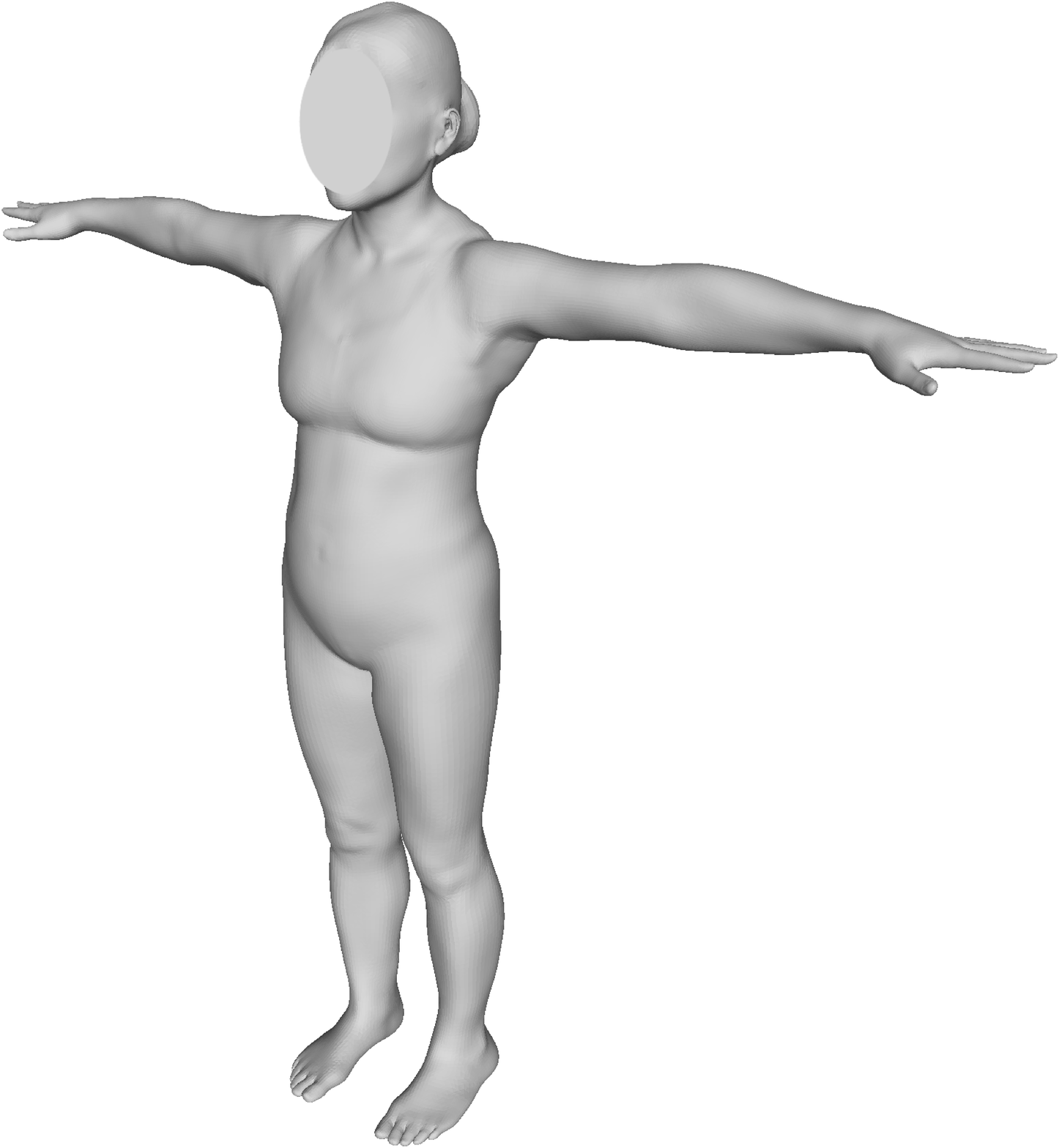}}
\hfill
    \subfigure[Predicted DXA scan imaging channels and corresponding error maps (represented as percentages) resulting from the 3D body scan (Figure~\ref{fig:3dscan}) and differing modeling methods.]{\label{fig:pseudo-error}
    \includegraphics[width=.64\textwidth]{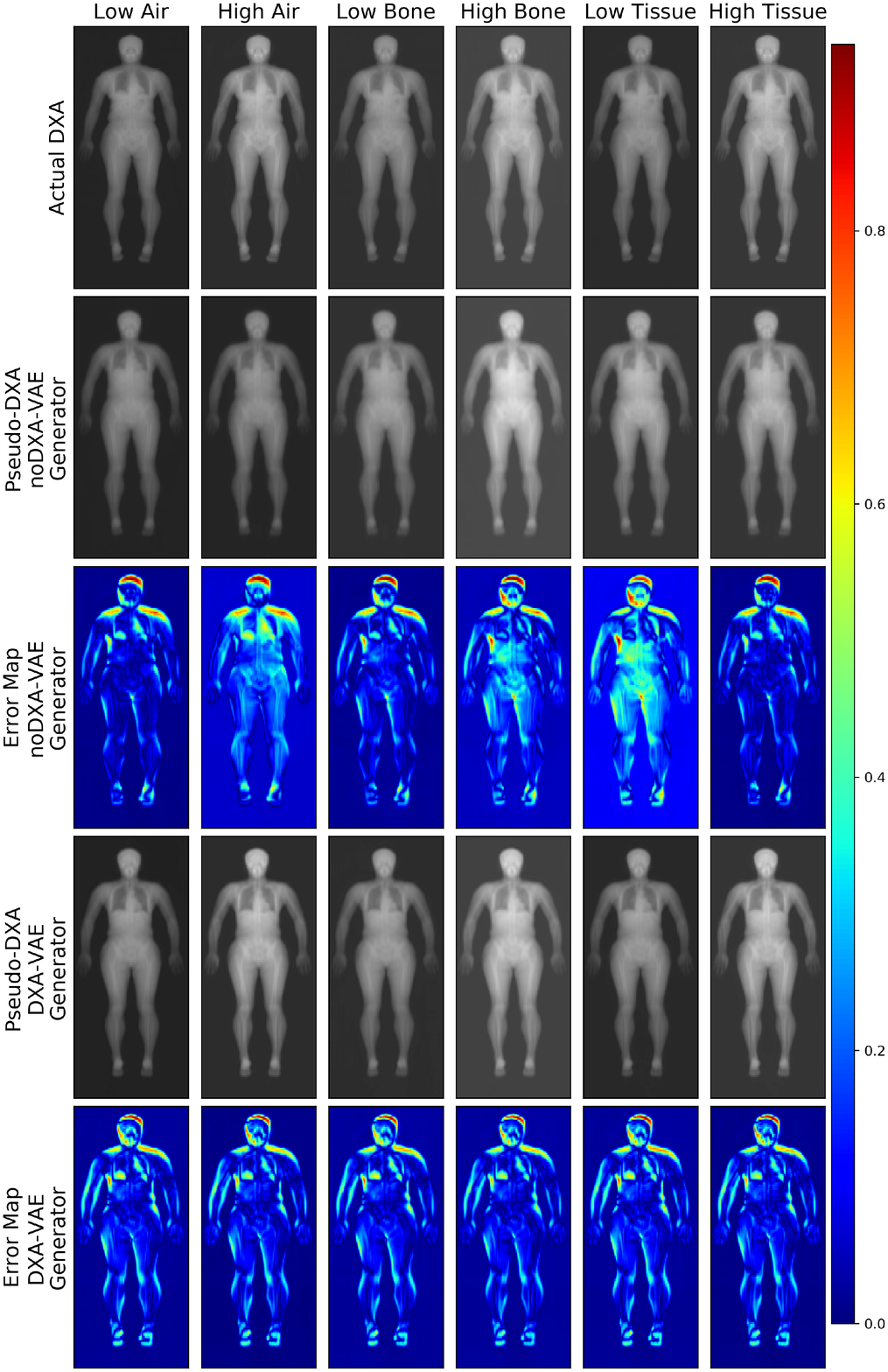}}
    \caption{Representative participant 3D scan and predicted DXA image with error map produced by Pseudo-DXA models in which generators were pretrained with and without the DXA loss.}
    \label{fig:pseudo-dxa}
\end{figure}

The Pseudo-DXA which incorporated the trained VAE generator yielded better image quality metrics when compared to the Pseudo-DXA models with randomly initialized weights. Figure~\ref{fig:pseudo-dxa} contains the input 3D body scan and Figure~\ref{fig:3dscan} contains the actual corresponding DXA image and predicted images from both models accompanied by their error maps. Figure~\ref{fig:pseudo-error} shows better DXA image predictions are produced from the model using the SSL VAE generator. The randomly initialized weights model fails to produce clear DXA images and large prediction errors can be seen in the low energy tissue channel. Positioning differences account for some of the significant errors (red and orange regions) visualized on the error maps in Figure~\ref{fig:pseudo-dxa}. Some image quality metrics are influenced by positioning difference however, body composition assessment is invariant to differences in position. 

\begin{table}[h!]
  \caption{Body composition analysis of DXA image predicted from 3D body surface by networks generators trained without (noDXA-VAE) and with (DXA-VAE) the DXA loss functions}
  \label{tab:pseudo-apex}
  \centering
\resizebox{.69\textwidth}{!}{
\begin{tabular}{lrrrr}
\toprule
Measurments & \multicolumn{1}{l}{\begin{tabular}[c]{@{}l@{}}\rsquared\\ noDXA-VAE \\  Generator\end{tabular}} & \multicolumn{1}{l}{\begin{tabular}[c]{@{}l@{}}\rsquared\\ DXA-VAE \\ Generator\end{tabular}} & \multicolumn{1}{l}{\begin{tabular}[c]{@{}l@{}}RMSE\\ noDXA-VAE\\ Generator\end{tabular}} & \multicolumn{1}{l}{\begin{tabular}[c]{@{}l@{}}RMSE\\ DXA-VAE\\ Generator\end{tabular}} \\
\midrule
\midrule
Arm BMC & 0.31 & \textbf{0.71} & 0.05 & 0.03 \\
Arm BMD & 0.17 & \textbf{0.67} & 0.14 & 0.08 \\
Arm Fat & 0.28 & \textbf{0.52} & 0.74 & 0.66 \\
Arm Lean & 0.40 & \textbf{0.74} & 1.18 & 0.65 \\
Leg BMC & 0.29 & \textbf{0.73} & 0.12 & 0.07 \\
Leg BMD & 0.06 & \textbf{0.52} & 0.22 & 0.15 \\
Leg Fat & 0.38 & \textbf{0.54} & 1.65 & 1.55 \\
Leg Lean & 0.31 & \textbf{0.79} & 2.68 & 1.51 \\
Total BMC & 0.20 & \textbf{0.66} & 0.60 & 0.37 \\
Total BMD & 0.02 & \textbf{0.41} & 0.23 & 0.14 \\
Total Fat & 0.34 & \textbf{0.68} & 10.05 & 7.44 \\
Total Lean & 0.39 & \textbf{0.80} & 1.18 & 8.58 \\
Total Mass & 0.53 & \textbf{0.90} & 12.99 & 5.28 \\
\bottomrule
\end{tabular}
}
\end{table}

We deduce from Table~\ref{tab:pseudo-apex} that the DXA-VAE resulted in a better generator which was used to construct a Pseudo-DXA model that can produce more biologically accurate images. Pseudo-DXA containing the trained generator out performed the model containing the noDXA-VAE generator on all body composition values. It should be noted that image quality and body composition were lower when comparing outputs from Pseudo-DXA models to outputs from the VAE models. This is likely caused by a relatively weak mapping between the 3D and 2D DXA space and overfitting due to the smaller paired dataset.

\section{Conclusion}

To our knowledge, this is the first instance in which a deep learning model can generate medical images on which subsequent quantitative imaging analysis can be performed accurately. In this work we frame medical images as quantitative maps rather than just anatomical representations. In doing so, we were able to introduce a simple DXA loss function which constrained SSL of a VAE to learn the specific pixel relationships between image channels. These specific relationships correspond to biological quantities of soft tissue and bone which allow for the derivation of body composition. Subnetworks of the trained DXA-VAE demonstrated superior performance when trained for downstream specific tasks which includes predicting 3D anthropology from DXA scans and producing real analyzable DXA scans from 3D body scans. The broader impact of this work is to demonstrate that medical image deep learning should consider the principles used to generate or produce images. This will lead towards more accurate synthesized or reconstructed images that can then be better integrated into clinical practice.

\iffalse %not included during submission time
\section*{Acknowledgments}
\label{sec:awk}
We thank all the participants and their parents for graciously giving us their time for all studies. The data underlying this study cannot be made publicly available because the data contain patient-identifying information. Data are available from the Shape Up! studies for researchers who meet the criteria for access to confidential data. For details and to request an application, visit https://shepherdresearchlab.org/.
\fi

\clearpage
\bibliography{ieetr}
\bibliographystyle{ieeetr}

\clearpage
%\input{appendix}

%%%%%%%%%%%%%%%%%%%%%%%%%%%%%%%%%%%%%%%%%%%%%%%%%%%%%%%%%%%%%%%%%%%%%%%%%%%%%%%%%%%

\end{document}